\documentclass[prl,showpacs,superscriptaddress,twocolumn,
floatfix,preprintnumbers,altaffilletter]{revtex4}
\usepackage{graphicx,url,amssymb,amsmath,longtable,rotating,color,units,wasysym,subfigure,wrapfig,epsfig}

\begin{document}

\title{Correlated magnetic noise in global networks of gravitational-wave interferometers: observations and implications}

\author{E. Thrane}
\address{LIGO-California Institute of Technology, Pasadena, California 91125, USA}
\email{ethrane@ligo.caltech.edu}
\author{N. Christensen}
\address{Physics and Astronomy, Carleton College, Northfield, MN 55057, USA}
\author{R. M. S. Schofield}
\address{University of Oregon, Eugene, Oregon 97403, USA}

\begin{abstract}
One of the most ambitious goals of gravitational-wave astronomy is to observe the stochastic gravitational-wave background.
Correlated noise in two or more detectors can introduce a systematic error, which limits the sensitivity of stochastic searches.
We report on measurements of correlated magnetic noise from Schumann resonances at the widely separated LIGO and Virgo detectors.
We investigate the effect of this noise on a global network of interferometers and derive a constraint on the allowable coupling of environmental magnetic fields to test mass motion in gravitational-wave detectors.
We find that while correlated noise from global electromagnetic fields could be safely ignored for initial LIGO stochastic searches, it could severely impact Advanced LIGO and third-generation detectors.  
\end{abstract}

\maketitle

{\em Introduction.}---One of the science goals of gravitational-wave (GW) astronomy is to measure the stochastic gravitational-wave background (SGWB).
In cosmological models, the SGWB can be created from inflationary physics~\cite{grishchuk,starob,eastherlim,maggiore}, cosmic strings~~\cite{caldwellallen,DV2}, and pre-Big-Bang models~\cite{PBB1,PBBpaper}.
Detection of the SGWB would constitute a remarkable discovery and could offer a unique probe of the history of the universe back to the earliest moments after the Big Bang~\cite{maggiore}.
The SGWB can also be created from the superposition of many astrophysical objects such as highly magnetized stars~\cite{cutler,RegMan}, young or spun-up neutron stars~\cite{RegPac,owen,barmodes1,barmodes3}, core collapse supernovae~\cite{firststars}, white dwarf binaries~\cite{phinney_whitedwarfs}, super-massive black hole binaries~\cite{jaffe}, and perhaps most promisingly, compact binary coalescences (CBCs)~\cite{phinney,kosenko,regimbau}.
Recent work~\cite{zhu,StochCBC} suggests that the SGWB from CBCs may be detectable with Advanced LIGO (aLIGO) and Advanced Virgo~\cite{aLIGO2,aVirgo}.

The standard procedure for searches for the SGWB is to cross-correlate strain data channels from two detectors~\cite{allen-romano,stoch-S5,sph_results,stoch-S4,radiometer}).
By integrating data obtained over a $\approx\unit[1]{yr}$-long run, it has been possible to achieve astrophysically interesting results.
The initial LIGO~\cite{iligo} and Virgo~\cite{Virgo} experiments, which took data from 2000-2010, yielded limits on the energy density of GWs that surpass indirect bounds from Big Bang Nucleosynthesis and the cosmic microwave background~\cite{stoch-S5}.
The aLIGO experiment~\cite{aLIGO2}, scheduled to begin taking data in 2015, is expected to improve on past results by a factor of $\approx10^4$~\cite{zhu,paramest}.

A key assumption underlying the SGWB search strategy is that the noise in each detector is uncorrelated~\cite{allen-romano,christensen,christensen_prd}.
This is most easily achieved in spatially separated interferometers, and previous analyses~\cite{stoch-S5,sph_results,s5vsr1,stoch-S4,radiometer} using the LIGO Hanford (LHO) and LIGO Livingston (LLO) observatories---separated by a distance of $\approx\unit[3000]{km}$---exhibited no evidence of correlated environmental noise.
(Self-inflicted correlated noise artifacts, such as a $\unit[1]{Hz}$ comb from similar electronics components at each site, have previously been identified and notched.)
As GW interferometers become more sensitive, however, it is possible for subtle global phenomena to produce correlated noise at problematic levels.

In the absence of correlated noise, the sensitivity of a search for the SGWB is limited only by the observation time $t_\text{obs}$, with the signal-to-noise ratio growing like $t_\text{obs}^{1/2}$.
Correlated noise, however, produces a systematic error, which cannot be reduced through integration.
To the extent that it cannot be mitigated through instrumental (re)design and/or background subtraction, it constitutes a fundamental limit for SGWB searches.

Previous work~\cite{allen-romano,christensen,christensen_prd} has identified Schumann resonances as a potential source of correlated noise for widely separated detectors, especially for second-generation experiments such as aLIGO.
Schumann resonances, predicted in 1952~\cite{schumann,schumann-b} and observed soon thereafter~\cite{schumann-c,balser-wagner}, are global electromagnetic (EM) resonances in the cavity formed by the surface of the Earth and the ionosphere.
The cavity is excited by a background of $\approx100$ lightning strikes per second around the world with $\unit[20-30]{kA}$ of current and lengths of $\unit[3-5]{km}$.
The resonances produce magnetic fields on the Earth's surface of $\unit[0.5-1.0]{pT \, Hz^{-1/2}}$~\cite{Sentman}.
Observed in the time domain, $\unit[10]{pT}$ bursts appear above a $\unit[1]{pT}$ background~\cite{fullekrug} at a rate of $\approx\unit[0.5]{Hz}$.
The primary resonance is at $\unit[8]{Hz}$ with secondary and tertiary harmonics at $\unit[14]{Hz}$ and $\unit[20]{Hz}$ respectively.
The peaks exhibit a spectral width of $\approx20\%$ and vary seasonally and with proximity to lightning storms.

Global magnetic fields such as Schumann resonances can cause correlated noise in gravitational-wave interferometers by inducing forces on magnets mounted directly on the test masses for position control (as in initial LIGO) or on magnets (and magnetically susceptible material) higher up the test-mass suspension system (as in aLIGO).
Since Schumann resonances produce coherent EM fields over $\approx\unit[1000]{km}$ distances, it may be possible for them to introduce correlated strains in  widely separated detectors.
Recent observations of Schumann resonances and a discussion of their properties are given in~\cite{Sentman,price,shvets}.
Other EM phenomena such as solar storms, currents in the van Allen belt~\cite{vanAllen}, and anthropogenic emission may also contribute to correlated EM noise.

In this Letter we report on measurements of correlated EM noise (including contributions from Schumann resonances) observed in magnetometers situated at widely separated GW observatories.
Combining these results with prior measurements of the coupling between magnetic fields and LIGO test mass motion, we infer the level of correlated strain noise present in initial LIGO SGWB analyses.
Then we derive magnetic isolation specifications for the next generation of advanced detectors to ensure that future SGWB analyses are not limited by correlated noise.
We conclude by discussing the implications for third-generation detectors and possible strategies for subtracting correlated noise.

{\em Formalism.}---Searches for the SGWB typically seek to measure the logarithmic energy density of GWs
\begin{equation}
  \Omega_\text{GW}(f)  = \frac{1}{\rho_c} \frac{d\rho_\text{GW}}{d\text{ln}f} ,
\end{equation}
where $f$ is GW frequency, $\rho_c$ is the critical energy density required for a closed universe, and $d\rho_\text{GW}$ is the energy density of GWs between $f$ and $f+df$.
An optimal estimator for $\Omega_\text{GW}$, integrated over some detection band, can be constructed from the strain time series of two GW interferometers~\cite{allen-romano}:
\begin{equation}\label{eq:Y}
  \hat{Y}(f) = \frac{2}{\delta T} \text{Re}\left( 
  Q(f) \tilde{s}_1^\star(f) \tilde{s}_2(f)
  \right).
\end{equation}
Here $Q(f)$ is a filter function (see~\cite{allen-romano}), $\delta T$ is the data segment duration, and $\tilde{s}_I(f)$ is the Fourier transform of the strain series measured by detector $I$.
By combining data from many data segments and frequency bins, it is possible to construct an optimal broadband estimator for the entire science run, (see~\cite{allen-romano} for details).

Following~\cite{allen-romano}, the strain observed by each detector contains contributions from signal and noise and can be written as
\begin{equation}
  \tilde{s}_I(f) = \tilde{h}_I(f) + \tilde{n}_I(f), 
\end{equation}
where $\tilde{h}_I(f)$ is the GW strain induced in detector $I$ and $\tilde{n}_I(f)$ is the noise.
If $\tilde{n}_1(f)$ and $\tilde{n}_2(f)$ are uncorrelated, an optimal filter $Q(f)$ can be chosen such that
\begin{equation}\label{eq:expectation}
  \langle \hat{Y}(f) \rangle = \Omega_\text{GW}(f) ,
\end{equation}
where the angled brackets denote an expectation value.
An estimator for the uncertainty of $\hat{Y}(f)$, denoted $\hat\sigma_Y(f)$, grows like $t_\text{obs}^{-1/2}$.
In the absence of correlated noise, $\hat{Y}(f)$ is an unbiased estimator and the sensitivity of the search is limited by $t_\text{obs}$.

In the presence of correlated noise, Eq.~\ref{eq:expectation} becomes
\begin{equation}
  \langle \hat{Y}(f) \rangle = \Omega_\text{GW}(f) + \Omega_N(f)
\end{equation}
where 
\begin{equation}
  \Omega_N \equiv \frac{2}{\delta T} \text{Re} \left[
    Q(f) \langle \tilde{n}^\star_1(f) \tilde{n}_2(f) \rangle
  \right]
\end{equation}
is the correlated noise.
$\Omega_N(f)$ represents a systematic bias.
If $\Omega_N(f)\gtrsim\hat\sigma_Y(f)$, then the search is limited by how well $\Omega_N(f)$ can be estimated.
If, however, $\Omega_N(f)\ll\hat\sigma_Y(f)$, then the correlated noise term may be safely ignored.

Before we continue, we define {\em coherence}, which is useful for determining if two channels are correlated.
The coherence between channels $1$ and $2$ is given by:
\begin{equation}\label{eq:coh}
  \text{coh}(f) \equiv 
  \frac{\left|\overline{\tilde{s}_1^\star(f) \tilde{s}_2(f)}\right|^2}
       {\overline{\left|\tilde{s}_1(f)\right|^2} \,
         \overline{\left|\tilde{s}_2(f)\right|^2} } .
\end{equation}
Here the overline denotes time-averaging over $N$ segments.
If $s_1$ and $s_2$ are independent, Gaussian, and stationary random variables, then $\langle\text{coh}(f)\rangle=1/N$.
Deviations from $\text{coh}(f)\approx1/N$ are evidence that one of these three assumptions is violated.
Integration over long observation periods probes below the uncorrelated noise to unearth low-level coherent features.

It is also useful to define coherence in terms of cross- and autopower
\begin{equation}\label{eq:coh2}
  \text{coh}(f) \equiv \frac{\left|\overline{S_{12}(f)}\right|^2}
       {\overline{S_1(f)} \, \overline{S_2(f)}}, 
\end{equation}
where
\begin{equation}
  S_I(f) \equiv \frac{1}{\cal N} \left|\tilde{s}_I(f)\right|^2
\end{equation}
is the autopower for detector $I$ and
\begin{equation}
  S_{12}(f) \equiv\frac{1}{\cal N} \tilde{s}_1^\star(f) \tilde{s}_2(f)
\end{equation}
is the cross-power for the detector pair $1$ and $2$.
${\cal N}$ is a discrete Fourier transform normalization constant.
We respectively refer to $\sqrt{S_I(f)}$ and $\sqrt{\left|S_{12}(f)\right|}$ as the auto- and cross- {\em amplitude} spectra.

\begin{figure}[hbtp!]
  \psfig{file=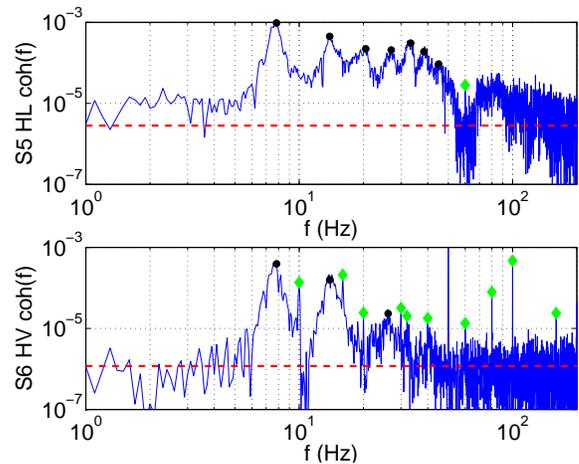,width=3in}
  \caption{
    Magnetometer coherence spectra for LHO-LLO during the LIGO S5 science run (top, $t_\text{obs}=\unit[330]{dy}$) and for LHO-Virgo during S6-VSR2/3 (bottom, $t_\text{obs}=\unit[100]{dy}$).
    Schumann resonance peaks are indicated with black circles while electronic noise lines are indicated with green diamonds.
    The red dashed line indicates the average value expected for uncorrelated noise.
    Some LHO-LLO peaks are obscured by $\unit[60]{Hz}$ electronic noise.
    The frequency resolution is $\unit[0.1]{Hz}$.
    \label{fig:coh_spec}
  }
\end{figure}

{\em Procedure and results.}---The LIGO and Virgo observatories are equipped with sensors to measure environmental noise that may affect strain measurements~\cite{iligo}.
Of particular interest here are LIGO's Bartington Mag-03 magnetometers and Virgo's Metronix MFS-06 magnetometers, both housed inside observatory buildings.
We calculate $\text{coh}(f)$ for magnetometer channels at different detectors to determine if they exhibit excess coherence.
Data are broken into $\unit[10]{s}$ segments yielding $\unit[0.1]{Hz}$ resolution.
Example multi-month coherence spectra are shown in Fig.~\ref{fig:coh_spec} for LHO-LLO and LHO-Virgo magnetometer pairs.
There are several noteworthy features.


First, we observe Schumann peaks (marked with black circles), three of which are coherent between all three observatories.
The peak frequencies $f=\unit[7.8, 13.9, 20.5]{Hz}$ etc.\ are consistent with expected values.
While Fig.~\ref{fig:coh_spec} shows just two spectra, we observe similar peaks for channel pairs associated with different magnetic field directions~\footnote{While boundary conditions dictate that the vertical component of magnetic Schumann fields is zero, we observe that the warping of field lines due to metal structures leads to significant Schumann fields in all directions.}, for magnetometers located at different locations at the observatories, and for data taken during different observing periods.
By plotting the cumulative coherence as a function of time, we determine that the observed peaks are not due to non-stationarity effects (a small number of data segments with anomalously high power).
We observe no broadband coherence at high frequencies $f\gtrsim\unit[200]{Hz}$ for LHO-LLO and $f>\unit[50]{Hz}$ for LIGO-Virgo.

Second, we observe narrowband features (marked with green diamonds) at frequencies that are expected for noise from electronics systems.
The LHO-LLO coherence spectrum exhibits an expected line at $\unit[60]{Hz}$ from the electrical power systems.
The LHO-Virgo spectrum exhibits lines at $\unit[16]{Hz}$, $\unit[50]{Hz}$ (extending beyond the plot range), and $\unit[100]{Hz}$.
The lines at $\unit[50]{Hz}$ and $\unit[100]{Hz}$ are produced by nearly monochromatic electrical power system frequencies at Virgo aligning in frequency with members of a $\unit[10]{Hz}$ comb produced by a timing synchronization system at LHO.
Since we are concerned here with magnetic fields that are correlated over global distance scales, we subsequently remove these line artifacts.

Finally, as an additional cross-check, we repeated the coherence calculation while shifting one time series $\unit[10]{s}$ with respect to the other.
The time-shifted spectra exhibit no broadband excess while electronic lines remain, which supports our conclusion that the broadband peaks are due to global magnetic correlations.

{\em Implications.}---In order to investigate the effect of EM correlated noise during initial LIGO, we utilize measurements of the coupling function $T(f)$ (the absolute value of the transfer function), which describes the induced test mass motion per unit environmental magnetic field.
The measurements, performed multiple times during each science run, utilize magnetic injection coils (placed outside but nearby the vacuum chambers) to create alternating magnetic fields.
The injection coils are located at a distance great enough that the magnetic field gradients at the test mass are dominated by the effects of local field-altering structures.
The resultant test mass motion is measured using the GW strain channel.
The initial LIGO magnetic coupling functions for each test mass at both sites were found to be within a factor of two of $T(f) = \unit[0.003 (f/\unit[1]{Hz})^{-3}]{m/T}$~\cite{robert}.
Also, the coupling function was found to be the same (to within a factor of two) for magnetic fields injected both perpendicular and parallel to the beam axis because of local field-altering structures~\cite{robert}.
We therefore proceed with the simplifying assumption that the coupling function is approximately the same at each interferometer: $T_1(f)=T_2(f)=T(f)$.
Further, in the analysis that follows, we conservatively use the magnetometer spectra exhibiting the strongest coherence at each LIGO site.
The least coherent spectra are approximately an order of magnitude less coherent, which would yield weaker coupling function requirements by $\approx 2$.

\begin{figure}[hbtp!]
  \psfig{file=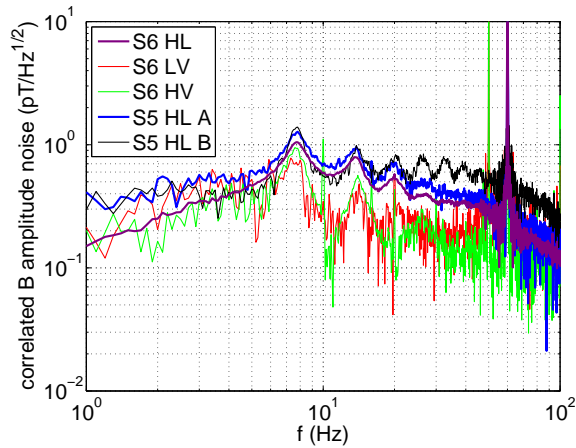,width=3in}
  \caption{
    Magnetic cross-amplitude spectra $\sqrt{\left|M_{12}(f)\right|}$ during the S5 and S6-VSR2/3 LIGO-Virgo science runs.
    Note that Fig.~\ref{fig:powerspec} and Fig.~\ref{fig:coh_spec} are related through Eq.~\ref{eq:coh2}.
    The frequency resolution is $\unit[0.1]{Hz}$.
    The blue S5 HL~A and violet S6 HL traces utilize the same magnetometer pair, whereas the black S5 HL~B trace comes from a magnetometer pair only available during S5.
  }
  \label{fig:powerspec}
\end{figure}

The strain noise cross-power spectrum induced from magnetic fields is 
\begin{equation}\label{eq:coupling}
  N^\text{m}_{12}(f) \approx \frac{1}{L^2} 
  T_1(f) T_2(f)\left| M_{12}(f) \right| ,
\end{equation}
where
\begin{equation}
  M_{12}(f)\equiv \frac{1}{\cal N}\overline{\tilde{m}_1^\star(f)\tilde{m}_2(f)}
\end{equation} 
is the magnetometer cross-power spectrum (see Fig.~\ref{fig:powerspec}), $\tilde{m}_I(f)$ is data from magnetometer channel $I$, and $L$ is the interferometer arm length.
By relating $N^\text{m}_{12}(f)$ to the absolute value of $M_{12}(f)$, we conservatively assume that the phase of $M_{12}(f)$ yields the worst possible contamination in the strain channel.
The superscript $m$ in Eq.~\ref{eq:coupling} labels this noise spectrum as correlated strain noise of magnetic origin as opposed to uncorrelated strain noise.

In the absence of correlated noise, the sensitivity of a stochastic search is determined by the uncorrelated noise spectrum, estimated by a weighted average of the product of the autopower spectra~\cite{allen-romano}:
\begin{equation}\label{eq:uncorrelated}
  N^\text{u}_{12}(f) = \frac{1}{N^{1/2}} 
  \left(\overline{S_1^{-1}(f) S_2^{-1}(f)}\right)^{-1/2} .
\end{equation}
The superscript $u$ in Eq.~\ref{eq:uncorrelated} labels this noise spectrum as uncorrelated noise, which decreases with $t_\text{obs}$.

\begin{figure}[hbtp!]
  \psfig{file=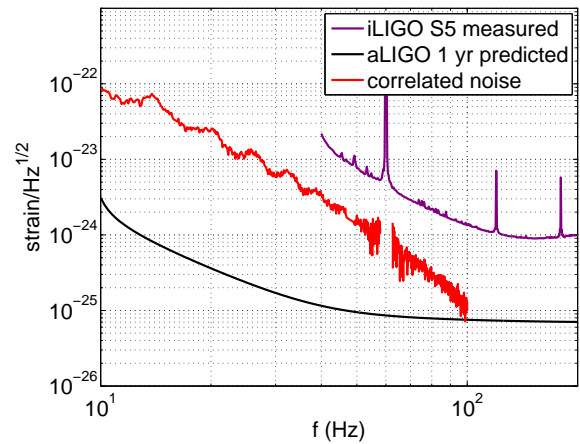,width=3in}
  \caption{
    Strain amplitude spectra for correlated and uncorrelated noise.
    Black is the uncorrelated noise $\sqrt{N^\text{u}_{12}(f)}$ for the H1L1 detector pair operating at Advanced LIGO design sensitivity and assuming $\unit[1]{yr}$ of integration.
    Purple indicates the uncorrelated noise $\sqrt{N^\text{u}_{12}(f)}$ achieved during initial LIGO using $\approx\unit[300]{dy}$ of coincident data.
    Red is $\sqrt{N^\text{m}_{12}(f)}$ (the estimated correlated noise due to EM fields during initial LIGO).
    Electronic noise lines have been notched.
    The spectra have been scaled to assume a frequency resolution of $\unit[0.25]{Hz}$, which is typical for stochastic searches~\cite{stoch-S5}.
  }
  \label{fig:noisebudget}
\end{figure}

In Fig.~\ref{fig:noisebudget} we show the correlated strain noise amplitude spectrum $\sqrt{N^\text{m}_{12}(f)}$ for the initial LIGO $\unit[4]{km}$ H1 and L1 detectors located at LHO and LLO respectively (red).
Alongside, we plot $\sqrt{N^\text{u}_{12}(f)}$ for aLIGO at design sensitivity with $\unit[1]{yr}$ of coincident data (black), and for initial LIGO with $\unit[330]{dy}$ of coincident data (purple).
We estimate that the correlated noise was below the uncorrelated noise during initial LIGO, which is consistent with back-of-the-envelope estimates from~\cite{allen-romano}, previous noise studies~\cite{vuk_shivaraj}, and past observational results~\cite{stoch-S5}.

\begin{figure}[hbtp!]
  \psfig{file=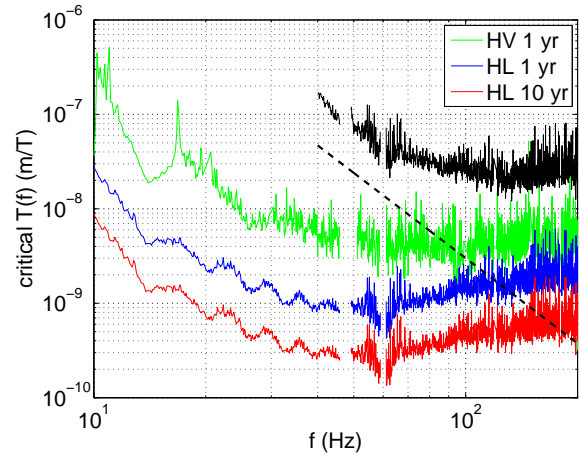,width=3in}
  \caption{
    The design-sensitivity Advanced Virgo and aLIGO critical coupling function $T(f)$ above which correlated noise is comparable to uncorrelated noise.
    The critical coupling function is calculated by considering the sensitivity obtained from a given pair of detectors, each assumed to have the same coupling function.
    Electronic noise lines have been notched.
    We also include the critical coupling function (solid black) for initial LIGO.
    The measured initial LIGO coupling function (dashed black) falls below it.
  }
  \label{fig:coupling}
\end{figure}

The aLIGO experiment, however, is expected to achieve a factor of $10$ better strain sensitivity than initial LIGO with a wider detection band going as low as $\unit[10]{Hz}$.
Using Eq.~\ref{eq:coupling}, and calculating $N^\text{u}_{12}(f)$ for aLIGO design sensitivity, we can solve for $T(f)$---the coupling function at which the correlated noise becomes comparable to the uncorrelated noise.
$T(f)$ is shown in Fig.~\ref{fig:coupling} for two different observation durations.
We also include $T(f)$ calculated for the LHO-Virgo detector pair.

The coupling function in Fig.~\ref{fig:coupling} may not be conservative enough because even when the correlated noise is below the uncorrelated noise, it can become significant through integration if it occurs in many frequency bins.
The enhancement of sub-threshold signals grows like the square root of the number of frequency bins.
Given a $\unit[100]{Hz}$ band of correlated noise and a resolution of $\unit[0.25]{Hz}$, it would be prudent to aim for a factor of $\sqrt{400}\approx20$ lower than the coupling shown in Fig.~\ref{fig:coupling}.

Correlated magnetic events may also be of concern for GW burst searches.
A $\unit[10]{pT}$ magnetic event and a coupling of $\unit[1\times 10^{-7}]{m/T}$ would generate a strain of $h \approx 2.5 \times 10^{-22}$.
However, the limited coherence $\text{coh}(f)\approx10^{-3}$ would likely preclude coincident false detections.
Magnetic bursts and their influence on the interferometers are the subject of ongoing research. 

The best means of guarding against EM-induced correlated noise is to minimize magnetic coupling to the test masses, e.g., through the removal of magnetic components.
As future interferometers, e.g.,~\cite{ET}, look to further reduce (uncorrelated) strain noise and to operate at lower frequencies, the required coupling function constraints will become increasingly stringent, and other remedies may be necessary.

One possibility worthy of study is to use magnetometers to subtract the correlated noise, e.g., with a Wiener filter scheme.
Similar tactics are under investigation to subtract Newtonian gravitational noise in GW interferometers~\cite{driggers}.
Effective subtraction requires precise measurement of global EM fields, which can lurk underneath an order of magnitude stronger local fields.
Future studies are necessary to determine the residual contamination expected from realistic background subtraction.

\begin{acknowledgments}
We gratefully acknowledge the use of environmental monitoring data from the Virgo and LIGO experiments.
We thank Alan Weinstein and Vuk Mandic for helpful comments.
ET is a member of the LIGO Laboratory, which is supported by funding from United States National Science Foundation.
LIGO was constructed by the California Institute of Technology and Massachusetts Institute of Technology with funding from the National Science Foundation and operates under cooperative agreement PHY-0757058.
NC and RS are supported by NSF grants PHY-1204371 and PHY-0855686 respectively.
This paper has been assigned LIGO document number LIGO-P1200167.
\end{acknowledgments}

\bibliography{noisebudget}

\end{document}